# Redistricting Compactness as Constrained Perimeter Minimization: Soap Bubble Theory and Discrete Approximation

**Mark B. Garman**

*Professor of Finance Emeritus, Haas School of Business, University of California, Berkeley*

## Abstract

We propose a mathematical framework for redistricting compactness grounded in the classical soap bubble problem: the partition of a planar region into $N$ subregions of prescribed population that minimizes total shared boundary length. This formulation, which we call the *soap bubble model*, yields a physically and mathematically motivated global compactness criterion that complements existing per-district compactness measures. We carefully distinguish two formulations: minimizing physical boundary length $L$ in geographic coordinates, and minimizing a density-weighted length $\tilde{L}$ in the coordinates produced by the Monge–Ampère density-flattening map. In the transformed coordinates, the first-order conditions for a minimizer are precisely the classical two-dimensional Plateau's laws — circular arc boundaries, 120° triple junctions, and perpendicular boundary contact. In geographic coordinates, the correct Euler–Lagrange equation is $\kappa_{ij}(s) = (P_i - P_j)d(s)$: arc curvature varies proportionally with local population density, so boundaries curve more tightly through dense areas and gently through sparse ones. Remarkably, the 120° junction condition and 90° boundary condition are unchanged by the density, as they follow from a force-balance argument that does not involve $d$. We adopt the physical-length objective $L$ as the primary redistricting criterion — it minimizes actual miles of district boundary and requires no coordinate transformation — and derive the corresponding density-modified volume-preserving mean curvature flow at the formal first-variation level. The $N$-state Potts model on the census block adjacency graph with population-weighted pressure updates provides a natural discrete implementation of the same variational principle. We define a multiplicative gerrymandering metric decomposed into tactical and systemic components and discuss practical computation, including grid-pinning effects in the discrete setting.

# 1. Introduction

## 1.1 The Compactness Problem in Redistricting

Every decade, following the United States Census, each state redraws its congressional districts. The constitutional requirement, established in *Wesberry v. Sanders* [W64], is one-person, one-vote: districts must contain, as nearly as practicable, equal populations. Beyond equal population, most state redistricting statutes require districts to be *compact* — a term that has resisted precise mathematical definition despite decades of litigation and scholarship.

The visual intuition is clear. A compact district is roughly circular or convex, with smooth boundaries following natural geography or administrative lines. A non-compact district is elongated, tentacled, or riddled with peninsulas and corridors. The challenge is making this intuition precise enough to serve as a legal standard and a measurement tool.

Several quantitative compactness measures have been proposed. The **Polsby–Popper score** [PP91] measures $4\pi \cdot \mathrm{Area}/\mathrm{Perimeter}^2$ for each district; a circle scores 1. The **Reock score** [R61] measures the ratio of a district's area to the area of its circumscribed circle. The **Schwartzberg score** [S66] measures the perimeter-to-circumference ratio of an equal-area circle. Each is computed district by district, without reference to global optimality, and each is ad hoc in functional form. There is no theorem establishing them as the correct measures of compactness.

A notable exception to the per-district pattern appears in the state constitutions of Iowa and Colorado, which mandate compactness measured at the plan level by comparing the total jurisdiction area to the sum of all district perimeters [DT24, fn.15]. Since jurisdiction area is constant across plans, this is equivalent to minimizing the sum of district perimeters — or equivalently, total interior boundary length. These provisions therefore point, as a matter of constitutional design, toward the same plan-level perimeter objective formalized in Problem 2.2, without the variational framework developed here to explain what the optimum configuration should look like. The present paper provides that foundation.

## 1.2 The Soap Bubble Connection

We propose an alternative foundation. The redistricting compactness problem is, in precise mathematical terms, a **constrained perimeter minimization problem**: partition the state into $N$ equal-population regions while minimizing the total length of the shared boundaries between regions. This is the same problem solved by soap bubbles.

When $N$ soap bubbles of specified volumes come into contact, the system minimizes total film length subject to fixed enclosed areas. The resulting configuration satisfies **Plateau's laws** [P73, T76, M94]. For the redistricting problem with *uniform* population density, these laws apply directly. For non-uniform density — the practically important case — one must take care to distinguish two formulations of the optimization, as explained in Section 3.1. This distinction is the mathematical core of the paper.

## 1.3 Relation to Existing Computational Approaches

**Integer programming** formulations [G70, V22] minimize total boundary length exactly but are intractable at census block resolution. **Markov chain Monte Carlo** methods, notably GerryChain [DDS21], sample the space of valid redistricting plans efficiently but optimize no specific criterion. **Weighted Voronoi / centroidal Voronoi tessellation** [Du99] produces clean districts with straight-line boundaries — the zero-curvature, zero-pressure-differential limit of our model.

Most directly related to the present work is the **cut score** of Duchin and Tenner [DT24], which counts the number of census block adjacency edges crossing district boundaries — the graph-theoretic analog of total boundary length. As shown in Section 4.2, the cut score is the special case of the Potts model energy with all edge weights equal to 1. The Potts model may thus be understood as the length-weighted generalization of the cut score, grounded in the variational principle of Problem 2.2. The cut score has been adopted in Maptitude redistricting software and cited in recent federal court proceedings in Alabama and Georgia.

Our approach identifies a physically motivated *optimality criterion* (minimize total boundary length), derives its *formal first-order structure* (density-modified Plateau's laws and the associated VPMCF), and motivates a discrete implementation through the constrained Potts model with population-weighted pressure. This model is GPU-parallel and scalable to full census block resolution, while questions of rigorous convergence for real census-block graphs are left as open numerical-analysis problems.

### 1.4 Main Contributions

1. **Formal variational framework.** Redistricting compactness as constrained perimeter minimization with non-uniform population density.
2. **Two objectives and the fork in the road.** Clear distinction between minimizing physical boundary length $L$ and transformed-coordinate boundary length $\tilde{L}$, with first-variation analysis of each.
3. **Plateau's laws for the transformed problem.** The standard result — circular arcs, 120° junctions, 90° boundary contact — holds for minimizers of $\tilde{L}$ in the Monge–Ampère transformed coordinates.
4. **Density-modified Plateau's laws for the physical problem.** For minimizers of $L$ in geographic coordinates, the correct Euler–Lagrange equation is $\kappa_{ij}(s) = (P_i - P_j)d(s)$: arc curvature proportional to local density. The 120° and 90° conditions are unchanged. Boundaries curve more tightly through dense areas — a compelling and testable prediction.
5. **Density-modified VPMCF.** At the formal level, the gradient flow for $L$ under the population constraint is $v_n = \kappa + (P_j - P_i)d(s)$, naturally consistent with the Potts model's population-weighted pressure.
6. **Potts model discretization.** The $N$-state Potts model on the block adjacency graph with population-weighted pressure is a natural discrete approximation motivated by the density-modified VPMCF.
7. **The gerrymandering metric.** A multiplicative decomposition into tactical and systemic components.

### 1.5 Organization

Section 2 states the variational problem, including the treatment of interior geographic features (lakes, rivers) as fixed domain boundaries. Section 3 derives the mathematical structure of minimizers for both objectives. Section 4 develops the discrete formulation, including the graph-level implementation of the geographic boundary treatment. Section 5 defines the gerrymandering metric. Section 6 discusses non-uniqueness and open problems.

---

# 2. The Variational Problem

## 2.1 Domain: States with Interior Geographic Features

Let $\Omega \subset \mathbb{R}^2$ denote the geographic domain of a state available for redistricting. In the simplest case $\Omega$ is simply connected, but in practice many states contain lakes, wide rivers, bays, and other uninhabited or unavailable regions that districts cannot include. We model these features as **interior exclusion zones**. Thus

$$\Omega = \Omega_0 \setminus (W_1 \cup \cdots \cup W_k),$$

where $\Omega_0$ is the outer state region and $W_1, \ldots, W_k$ are compact excluded sets. For the smooth variational formulation below, assume that $\Omega$ is a bounded, connected domain whose boundary consists of finitely many $C^2$ closed curves:

$$\partial\Omega = \partial\Omega_0 \cup \partial W_1 \cup \cdots \cup \partial W_k.$$

All components of $\partial\Omega$ are treated identically: they are fixed walls of the container. A district boundary arc that terminates on any component of $\partial\Omega$, whether the outer state boundary or a lake shore, satisfies the same natural orthogonality condition derived in Section 3.

This convention has an important modeling consequence. The length of $\partial\Omega$ contributes nothing to the objective $L$: the functional counts only interfaces separating two distinct districts, not arcs that lie on a fixed wall. A plan that places a lake shore, coastline, or excluded-zone boundary along the edge of a district incurs no additional interior-boundary penalty. In the discrete implementation of Section 4, the same convention is implemented by excluding water-area blocks and by assigning zero cost to appropriate natural-boundary edges.

Let $d : \overline{\Omega} \to \mathbb{R}_{>0}$ be a population density satisfying

$$d \in C^1(\overline{\Omega}), \qquad 0 < d_{\min} \le d(x) \le d_{\max} < \infty,$$

and let

$$P = \int_{\Omega} d \, dA$$

denote the total population. Let $N \ge 2$ be the number of congressional districts fixed by federal apportionment.

## Definition 2.1 (Population-Balanced Smooth Cluster)

A **population-balanced smooth $N$-cluster** in $\Omega$ is a collection

$$\mathcal{P} = \{\Omega_1, \ldots, \Omega_N\}$$

of pairwise disjoint open connected subsets of $\Omega$ satisfying the following conditions.

1. The closures cover the state up to sets of area zero:

$$\overline{\Omega} = \bigcup_{i=1}^{N} \overline{\Omega_i}.$$

2. Each district has equal population:

$$\int_{\Omega_i} d \, dA = \frac{P}{N}, \qquad i = 1, \ldots, N.$$

3. For every pair $i \ne j$, the district-district interface

$$\Gamma_{ij} = \partial\Omega_i \cap \partial\Omega_j \cap \Omega$$

   is a finite union of $C^2$ arcs.
4. The collection of interface arcs contains only finitely many junction points.

The **physical interior-boundary length** of the cluster is

$$L(\mathcal{P}) = \sum_{1 \le i < j \le N} \mathcal{H}^1(\Gamma_{ij}),$$

where $\mathcal{H}^1$ denotes one-dimensional Hausdorff measure. Only district-district interfaces contribute to $L$. Portions of district boundaries lying on the fixed state boundary, lake shores, coastlines, or other exclusion-zone boundaries are not counted.

## Problem 2.2 (Soap-Bubble Redistricting)

Among all population-balanced smooth $N$-clusters in $\Omega$, minimize

$$L(\mathcal{P}).$$

Equivalently, the soap-bubble redistricting problem asks for the partition of the state into $N$ connected equal-population districts having minimum total shared boundary length.

## Proposition 2.3 (Existence and Regularity Framework)

Under the assumptions above, and more generally in the standard finite-perimeter formulation of cluster minimization, the constrained perimeter-minimization problem admits a minimizing partition.

Furthermore, minimizers possess the classical local regularity structure of planar soap-bubble clusters: away from a finite set of singular points, district interfaces are smooth curves, and the local first-variation analysis applies. Interior singularities have the Plateau-type structure familiar from planar minimizing clusters, and free interfaces meeting the fixed container boundary satisfy the natural boundary condition associated with the perimeter functional.

The present paper does not seek the weakest possible hypotheses for existence or regularity. We use the classical theory of perimeter-minimizing clusters, in the sense of Almgren [A76] and Morgan [M94, M09], as the mathematical foundation for the variational analysis developed below.

*Discussion.* The purpose of Problem 2.2 is not to advance geometric measure theory, but to identify the physically motivated compactness objective for redistricting: minimizing actual district-to-district boundary length subject to equal population. Once that objective is specified, the first-order conditions for optimality lead directly to the density-modified Plateau laws derived in Section 3.

---

# 3. Mathematical Analysis of Minimizers

## 3.1 Two Objectives and the Fork in the Road

Because the population density $d$ is non-uniform, there are two natural formulations of the compactness problem, which in general have different minimizers.

**Path 1 (Physical length).** Minimize $L(\mathcal{P}) = \sum_{i<j} \mathcal{H}^1(\Gamma_{ij})$ — total boundary miles in geographic coordinates — subject to $\int_{\Omega_i} d\, dA = P/N$.

**Path 2 (Transformed length).** Let $\varphi : \Omega \to \tilde{\Omega}$ be the Brenier map (Section 3.4) that flattens $d$ to a uniform density $C$ on a target domain $\tilde{\Omega}$. Minimize $\tilde{L}(\mathcal{P}) = \sum_{i<j} \mathcal{H}^1(\varphi(\Gamma_{ij}))$ — arc length in the transformed coordinates — subject to equal *area* in $\tilde{\Omega}$ (which is equivalent to equal population in $\Omega$, by construction of $\varphi$).

When $d$ is constant, $\varphi$ is (up to scaling) the identity, and Path 1 and Path 2 coincide. In general they differ, with the discrepancy controlled by the spatial variation of $d$.

We take **Path 1 as our primary objective** because it minimizes the geographically interpretable quantity — actual miles of district boundary — and because, as Section 4 shows, its gradient descent is naturally implemented by the Potts model on the block graph without any coordinate transformation. Path 2 is developed first because it admits the cleaner classical analysis; its results then serve as the theoretical anchor for understanding the structure of Path 1 solutions.

## 3.2 Plateau's Laws for the Transformed Problem (Path 2)

In the target domain $\tilde{\Omega}$ with uniform density $C$, the minimization of $\tilde{L}$ subject to equal-area constraints is exactly the classical two-dimensional soap bubble problem. The standard first-variation analysis yields:

**Theorem 3.1 (2D Plateau's Laws in $\tilde{\Omega}$).** *Any smooth local minimizer of $\tilde{L}$ satisfies:*

*(i)* ***Constant curvature.*** *Each arc $\tilde{\Gamma}_{ij}$ in $\tilde{\Omega}$ is a circular arc.*

*(ii) $120^\circ$* ***junction condition.*** *At every interior junction, three arcs meet at $120^\circ$.*

*(iii) $90^\circ$* ***boundary condition.*** *Each arc meets $\partial\tilde{\Omega}$ perpendicularly.*

*Proof.* Under a normal perturbation $\tilde{\phi}(\tilde{s})$ of $\tilde{\Gamma}_{ij}$ in $\tilde{\Omega}$:
$\delta\tilde{L} = -\int_{\tilde{\Gamma}_{ij}} \tilde{\kappa}_{ij}(\tilde{s})\, \tilde{\phi}(\tilde{s})\, d\tilde{s}$.
Because the density in $\tilde{\Omega}$ is uniform ($C$), the variation of enclosed area is $\int \tilde{\phi}\, d\tilde{s}$. With Lagrange multipliers $\tilde{P}_i$, stationarity requires $\tilde{\kappa}_{ij} = \tilde{P}_i - \tilde{P}_j = \mathrm{const}$, establishing (i).

For (ii), at a triple junction where arcs $\tilde{\Gamma}_{ij}, \tilde{\Gamma}_{jk}, \tilde{\Gamma}_{ki}$ meet at $\tilde{\mathbf{x}}_0$, displacement of the junction point $\delta\tilde{\mathbf{x}}_0$ changes $\tilde{L}$ by $(\mathbf{t}_{ij} + \mathbf{t}_{jk} + \mathbf{t}_{ki}) \cdot \delta\tilde{\mathbf{x}}_0$. The area perturbation is second-order. Stationarity requires $\mathbf{t}_{ij} + \mathbf{t}_{jk} + \mathbf{t}_{ki} = \mathbf{0}$; three unit vectors summing to zero must meet at $120^\circ$.

For (iii), the natural boundary condition at a free contact point on $\partial\tilde{\Omega}$ forces perpendicularity. □

The Young–Laplace relation for Path 2 is $\tilde{\kappa}_{ij} = \tilde{P}_i - \tilde{P}_j$: the curvature of each arc equals the pressure difference between its flanking districts, where pressures are the Lagrange multipliers for the equal-area constraints in $\tilde{\Omega}$.

## 3.3 Density-Modified Plateau's Laws for the Physical Problem (Path 1)

We now derive the first-order conditions for smooth local minimizers of $L$ in physical coordinates. This is the central variational calculation of the paper. The distinction from the transformed problem is simple but important: length is still ordinary Euclidean length in geographic coordinates, while the constraint is population rather than area. Consequently, the first variation of the constraint carries the local density $d$.

Throughout this subsection we assume that the cluster is smooth in the sense of Definition 2.1, that $d \in C^1(\overline{\Omega})$, and that $d \geq d_{\min} > 0$. Let $\Gamma_{ij}$ be an oriented interface between districts $i$ and $j$, and let $\mathbf{n}_{ij}$ denote the unit normal pointing from $\Omega_i$ into $\Omega_j$.

**Lemma 3.2 (Variation of the population constraint).** *Let $\Gamma_{ij}$ be perturbed normally by $\phi(s)\mathbf{n}_{ij}$, where $s$ is arclength and $\phi$ is smooth and compactly supported away from junctions and endpoints. Then, to first order,*

$$\delta\left(\int_{\Omega_i} d\, dA\right) = -\int_{\Gamma_{ij}} d(s)\phi(s)\, ds, \qquad \delta\left(\int_{\Omega_j} d\, dA\right) = +\int_{\Gamma_{ij}} d(s)\phi(s)\, ds.$$

*Here $d(s) = d(\mathbf{x}(s))$ is the population density evaluated along the interface.*

*Proof.* A normal displacement by $\phi(s)\mathbf{n}_{ij}$ transfers, to first order, a strip of area $\phi(s)ds$ from $\Omega_i$ to $\Omega_j$. Integrating this infinitesimal transferred area against the continuous density $d$ gives the stated formula. Terms involving variation of $d$ across the strip are of second order in the displacement. □

**Theorem 3.3 (Density-Modified Plateau Laws in $\Omega$).** *Let $\mathcal{P}$ be a smooth local minimizer of $L$ among population-balanced smooth $N$-clusters. Then the following first-order conditions hold.*

*(i)* ***Density-proportional curvature.*** *Along each smooth interior interface $\Gamma_{ij}$,*

$$\kappa_{ij}(s) = (P_i - P_j)d(s),$$

*where $P_i$ and $P_j$ are the Lagrange multipliers associated with the population constraints of districts $i$ and $j$. Equivalently,*

$$\frac{\kappa_{ij}(s)}{d(s)} = P_i - P_j$$

*is constant along the interface.*

*(ii)* ***Triple-junction condition.*** *At every interior junction, three arcs meet at $120^\circ$.*

*(iii)* ***Boundary condition.*** *At a free endpoint on $\partial\Omega$, an interface meets the fixed boundary orthogonally.*

*Proof.* Consider first a normal variation supported in the interior of a single interface $\Gamma_{ij}$, away from junctions and endpoints. The first variation of arclength is

$$\delta\mathcal{H}^1(\Gamma_{ij}) = -\int_{\Gamma_{ij}} \kappa_{ij}(s)\phi(s)\,ds,$$

with the sign convention that positive $\phi$ moves the interface in the direction $\mathbf{n}_{ij}$. Introduce Lagrange multipliers $P_1, \ldots, P_N$ for the population constraints. The first variation of the constrained functional along $\Gamma_{ij}$ is therefore

$$\delta L - \sum_{m=1}^{N} P_m\,\delta\Big(\int_{\Omega_m} d\,dA\Big) = \int_{\Gamma_{ij}} [-\kappa_{ij}(s) + (P_i - P_j)d(s)]\phi(s)\,ds.$$

Stationarity for all such test functions $\phi$ gives

$$\kappa_{ij}(s) = (P_i - P_j)d(s),$$

which proves (i).

For a variation that moves an interior triple junction while keeping the adjacent arcs otherwise fixed to first order, the first variation of length is the vector force-balance term

$$(\mathbf{t}_{ij} + \mathbf{t}_{jk} + \mathbf{t}_{ki}) \cdot \delta\mathbf{x}_0,$$

where the $\mathbf{t}$'s are the unit tangent vectors of the three incident arcs, oriented away from the junction. The corresponding population changes are second order in the junction displacement, since the swept regions have area of order $|\delta\mathbf{x}_0|^2$. Thus the density-weighted constraints do not contribute to the first-order junction balance. Stationarity for arbitrary $\delta\mathbf{x}_0$ requires

$$\mathbf{t}_{ij} + \mathbf{t}_{jk} + \mathbf{t}_{ki} = 0.$$

Three unit vectors sum to zero exactly when they are separated by angles of $120^\circ$, proving (ii).

Finally, at a free endpoint on the fixed boundary $\partial\Omega$, the admissible endpoint displacement is tangent to $\partial\Omega$. The endpoint contribution to the first variation of length is the projection of the interface tangent onto this admissible boundary displacement. Since the population change at the endpoint is again second order, stationarity requires the interface tangent to be normal to $\partial\Omega$, equivalently the interface meets $\partial\Omega$ orthogonally. This proves (iii). ☐

**Remark 3.4 (Density-modified Young--Laplace relation).** The condition

$$\kappa_{ij}(s) = (P_i - P_j)d(s)$$

is the density-modified Young--Laplace relation for the physical redistricting problem. It reduces to the classical constant-curvature condition when $d$ is constant. Thus circular arcs are recovered in the uniform-density limit, while variable density produces variable curvature along an interface.

**Remark 3.5 (Geometric interpretation).** The equilibrium condition predicts that, for a fixed pressure difference $P_i - P_j$, boundaries have larger curvature in densely populated regions and smaller curvature in sparsely populated regions. In this sense, the optimal physical-length boundary responds to population density without changing the surface tension: the density enters through the constraint, not through the length functional. This is a testable prediction of the model in states with large urban-rural density contrasts.

**Remark 3.6 (What is preserved and what changes).** Passing from the transformed problem to the physical problem changes the arc-shape equation but not the angular balance laws. Circular arcs become density-weighted curves satisfying $\kappa/d = \mathrm{constant}$ along each district-district interface, while triple junctions still meet at $120^\circ$ and free boundary contacts remain orthogonal. These angular conditions are consequences of equal line tension and are independent of the population density.

## 3.4 The Density Transformation (Path 2 in Detail)

**Definition 3.7 (Brenier map).** The *Brenier map* $\varphi = \nabla\psi : \Omega \to \tilde{\Omega}$ is the gradient of a convex function $\psi$ satisfying the Monge–Ampère equation
$\det(D^2\psi(x,y)) = \frac{d(x,y)}{C}, \qquad \nabla\psi(\Omega) = \tilde{\Omega},$
where $C = P/|\tilde{\Omega}|$. Brenier's theorem [B91] guarantees existence and uniqueness of such a map.

Under $\varphi$, the population measure $d\,dA$ pushes forward to the uniform measure $C\,dA$ on $\tilde{\Omega}$, so equal-population districts in $(\Omega, d)$ correspond exactly to equal-area districts in $(\tilde{\Omega}, C)$. The minimizers of $\tilde{L}$ in $\tilde{\Omega}$ (Path 2) are the arcs satisfying Theorem 3.1; their preimages under $\varphi^{-1}$ are smooth curves in $\Omega$ whose curvature is determined by the Jacobian of $\varphi^{-1}$, and which are distinct in general from the minimizers of $L$ in $\Omega$ (Path 1).

**Relationship between the two optima.** Let $\mathcal{P}_L^*$ and $\mathcal{P}_{\tilde{L}}^*$ denote the (approximate) global minimizers of $L$ and $\tilde{L}$ respectively. When $d$ is nearly uniform, $\varphi$ is nearly the identity and the two optima nearly coincide. Quantifying the gap $|L(\mathcal{P}_L^*) - L(\mathcal{P}_{\tilde{L}}^*)|$ as a function of the spatial variation of $d$ — e.g., in terms of the Lipschitz constant of $\log d$ — is a natural open problem (Section 6.4). In states with large urban-rural density contrasts this gap may be non-negligible, and both solutions may be of interest: $\mathcal{P}_L^*$ minimizes physical boundary miles, while $\mathcal{P}_{\tilde{L}}^*$ minimizes a density-weighted boundary that gives more weight to boundaries in populated areas.

For the gerrymandering metric of Section 5 we adopt $L$ (Path 1), because it measures the geographically interpretable quantity and because the Brenier map is not needed computationally: population density enters the discrete algorithm only through the pressure term (Section 4.2).

## 3.5 Formal Gradient Flow: Density-Modified VPMCF

The same first-variation calculation suggests a natural constrained curvature flow. This subsection is meant in the formal classical sense: it describes the smooth gradient descent associated with $L$ before singularities, topology changes, or weak formulations are introduced. Establishing long-time existence and convergence for variable density and multiple districts is left as an open problem.

**Definition 3.8 (Formal density-modified VPMCF).** The formal volume-preserving mean-curvature flow associated with the physical-length objective is

$$v_n = \kappa_{ij}(s) + \big(P_j(t) - P_i(t)\big)d(s) \qquad \text{on } \Gamma_{ij}(t),$$

where $v_n$ is the normal velocity in the direction from $\Omega_i(t)$ to $\Omega_j(t)$. The time-dependent multipliers $P_i(t)$ are chosen so that

$$\frac{d}{dt}\int_{\Omega_i(t)} d\,dA = 0, \qquad i = 1, \dots, N.$$

This differs from standard uniform-density VPMCF by the factor $d(s)$ multiplying the pressure differential. The factor appears because a normal displacement through a dense region transfers more population than the same displacement through a sparse region.

**Proposition 3.9 (Formal dissipation identity).** *For a smooth solution of the formal density-modified VPMCF, before singularities or topology changes occur, the physical boundary length is nonincreasing:*

$$\frac{dL}{dt} \le 0.$$

*Proof.* The normal velocity is chosen as the negative constrained first variation of $L$ under the equal-population constraints. Equivalently, along each interface,

$$v_n = \kappa_{ij} - (P_i - P_j)d.$$

Using the standard first variation of length and the constraint equations, the derivative of $L$ along the projected gradient flow is

$$\frac{dL}{dt} = -\sum_{i<j}\int_{\Gamma_{ij}(t)} v_n^2\,ds \le 0.$$

The derivative vanishes precisely when $v_n = 0$ on every interface, which is the density-modified Young--Laplace condition of Theorem 3.3(i). ☐

**Corollary 3.10 (Formal fixed points).** *Smooth fixed points of the formal density-modified VPMCF satisfy the first-order conditions of Theorem 3.3. Conversely, any smooth cluster satisfying those first-order conditions is stationary for the formal flow.*

**Connection to the Potts model.** In the discrete formulation below, the effective energy change for reassigning block $v$ from district $d$ to district $d'$ has the form

$$\Delta E_{\text{eff}} = \Delta E_{\text{tension}} + p_v(P_d - P_{d'}).$$

The pressure-work term scales with block population $p_v \approx d(v)\,\mathrm{area}(v)$, mirroring the density factor in the continuum first variation. Thus the Potts model with population-weighted pressure is a natural discrete implementation of the same constrained perimeter principle. A proof that this dynamics converges to the density-modified VPMCF for irregular census-block graphs is not claimed here; it is identified in Section 6 as an open numerical-analysis problem.

---

# 4. Discrete Formulation

## 4.1 The Block Adjacency Graph

**Definition 4.1.** The *block adjacency graph* $G = (V, E, w, p)$ has: one node per census block; edges between blocks sharing a positive-length boundary; edge weight $w_{uv}$ equal to the physical length (meters) of the shared segment; node weight $p_v$ equal to total block population.

The **discrete boundary length**

$E(f) = \sum_{(u,v) \in E} w_{uv} \cdot \mathbf{1}[f(u) \neq f(v)]$

is the graph-cut analogue of the continuum interface length $L(\mathcal{P})$. For regular grids, random geometric graphs, and related point-cloud discretizations, graph total variation and discrete perimeter functionals are known to converge, under appropriate scaling hypotheses, to continuum total variation or perimeter functionals in the sense of $\Gamma$-convergence [CGL10, GTS16]. We use this literature as justification for treating $E(f)$ as a consistent discrete surrogate for physical boundary length, while recognizing that a full $\Gamma$-convergence theorem for the highly irregular census-block graph is a separate technical problem.

**Treatment of interior geographic features.** The multiply-connected domain structure of Section 2.1 is implemented in the graph as follows. Census blocks are classified by the TIGER/Line feature class codes into two categories:

- *Land blocks* ($p_v \geq 0$): standard redistributable blocks, forming the node set $V$.
- *Water-area blocks*: blocks classified as open water (lakes, wide rivers, bays), identified by TIGER feature class code `H` (hydrographic area). These are **excluded from** $V$ entirely; they correspond to the exclusion zones $W_1, \ldots, W_k$ of Section 2.1.

Edges in the graph are similarly classified:

- *Interior edges*: edges between two land blocks in different districts. These contribute $w_{uv}$ to $E(f)$ and are the targets of the optimization.
- *Shoreline edges*: edges between a land block and an excluded water block. These correspond to arcs on $\partial\Omega$ (lake shores, river banks). They are **excluded from** $E$ and contribute zero to $E(f)$, implementing the theoretical condition that $\partial\Omega$ contributes nothing to $L$.
- *Natural boundary edges*: edges between two land blocks that are separated by a linear water feature (a narrow river or stream recorded as a line, not an area, in TIGER). These are identified by the TIGER linear hydrography layer and assigned $w_{uv} = 0$, making it costless to place a district boundary along a natural waterway.

This classification implements the correct geometric treatment: using a lake shore or river as a district boundary is not penalized, since such boundaries are part of $\partial\Omega$ and do not enter the objective. In the continuum model, free interfaces meeting such fixed walls satisfy the $90^\circ$ perpendicularity condition of Theorem 3.3(iii); in the graph model this condition is represented only approximately at the resolution of the block geometry.

**Remark 4.2 (Michigan and multiply-connected states).** States with complex water topology — Michigan (two peninsulas, thousands of inland lakes), Minnesota (Land of Ten Thousand Lakes), Florida (coastline, the Everglades) — have correspondingly complex $\partial\Omega$ with many interior components. The framework handles these without modification: each lake shore is an additional fixed wall, district boundaries that reach the shore meet it perpendicularly, and the lake perimeter contributes nothing to $L$. These states provide the richest geometric test cases for the framework.

## 4.2 The Potts Model Energy and Pressure

The unconstrained problem is the ferromagnetic $N$-state **Potts model** on $G$ [Wu82]. Population balance is enforced by district pressures:
$P_i = \alpha\left(\sum_{v:\, f(v)=i} p_v - \frac{P}{N}\right), \qquad \alpha = \frac{\bar{w}}{\bar{p}},$
and the effective energy change for reassigning block $v$ from $d$ to $d'$ is:

$$\Delta E_{\text{eff}}(v, d \to d') = \underbrace{\sum_{\substack{u \in \mathcal{N}(v) \\ f(u)=d}} w_{vu} - \sum_{\substack{u \in \mathcal{N}(v) \\ f(u)=d'}} w_{vu}}_{\Delta E_{\text{tension}}} + \underbrace{p_v(P_d - P_{d'})}_{\text{pressure work}}.$$

As noted in Section 3.5, the pressure work scales with $p_v$, naturally implementing the density factor $d(s)$ of the density-modified VPMCF.

*Note on $\alpha$ in heterogeneous states.* The global choice $\alpha = \bar{w}/\bar{p}$ balances the tension and pressure terms at the state-wide average scale. In states with extreme density heterogeneity — where urban block populations may exceed rural block populations by two or three orders of magnitude — a spatially adaptive $\alpha(v)$ that varies with local density may ultimately be needed to prevent the pressure term from dominating in dense urban clusters while remaining negligible in deep rural areas; this refinement is left for future implementation work.

**Remark 4.3 (Relation to the cut score).** Setting $w_{uv} \equiv 1$ for all edges reduces $E(f)$ to the cut score of Duchin and Tenner [DT24], which counts the number of adjacency edges crossing district boundaries. The cut score is thus the degenerate case of the Potts energy in which physical boundary lengths are ignored and all block adjacencies are treated as equivalent. The length-weighted Potts energy preserves the coordinate-independence and resolution-stability motivation of the cut score while additionally encoding the physical boundary cost of each cut. Its continuum interpretation is supported by the discrete-to-continuum literature on graph total variation and discrete perimeters [CGL10, GTS16], although proving a census-block-specific convergence result is beyond the scope of this paper.

## 4.3 Contiguity Enforcement

Before accepting a move of $v$ from $d$ to $d'$, verify that $d$ remains connected after removing $v$. Let $\mathcal{N}_d(v) = \{u \in \mathcal{N}(v) : f(u) = d\}$. The move is *contiguity-safe* if $|\mathcal{N}_d(v)| \leq 1$ (trivially safe) or $\mathcal{N}_d(v)$ is connected in $G[d \setminus \{v\}]$ (verified by BFS, average cost $O(\deg(v)^2)$).

## 4.4 Grid Pinning and Mitigation

A limitation of gradient descent ($T = 0$) on the census block graph merits explicit discussion. Census blocks are predominantly rectilinear — bounded by street segments at roughly 90° angles. When the theoretically optimal boundary arc runs at an oblique angle to the local street grid, no single block reassignment may decrease $E$, even though the current configuration is geometrically inefficient. This **grid-pinning** effect means that $T = 0$ descent may halt at artificial local minima that are not genuine fixed points of the density-modified VPMCF.

Two mitigations are recommended:

*(i) Low-temperature annealing for $L_{\text{local}}$ estimation.* Rather than strict $T = 0$, use a very low but non-zero temperature $T_{\text{local}} \ll \bar{w}$ that allows the algorithm to cross small grid-pinning barriers while remaining effectively deterministic at the scale of the genuine local minimum. The local minimum is identified as the configuration at which no single-block move has $\Delta E_{\text{eff}} < -T_{\text{local}}$.

*(ii) Multi-block cluster swaps.* Augment the single-block update with occasional moves that reassign small contiguous clusters of 2–4 blocks simultaneously. Cluster swaps allow boundary segments to cross grid lines — for example, a 90° staircase can be replaced by a diagonal boundary in a single cluster-swap step. The contiguity check extends naturally to cluster moves.

Grid pinning does not affect the global-minimum estimation (Section 5), since the simulated annealing algorithm at high temperature already crosses all such barriers.

## 4.5 The 120° Junction in the Discrete Setting

The 120° junction condition of Theorem 3.3(ii) is a continuum property. In the discrete setting, census blocks meet at arbitrary angles — commonly 90° at street intersections — and an exact 120° balance cannot be realized. The discrete counterpart is a *locally stable triple boundary meeting*: a configuration in which no single-block move near the junction decreases $\Delta E_{\text{eff}}$. The junction "distributes" its tension over a small cluster of blocks, with the $120^\circ$ condition satisfied approximately at the scale of several blocks. Post-convergence verification should measure angles between boundary curves at junctions at the scale of 5–10 blocks (smoothing over block-level discretization), with typical tolerances of $\pm 5^\circ$–$10^\circ$.

## 4.6 Parallel Implementation

The local update is inherently parallel. **Graph coloring** (partition $V$ into color classes $C_1, \ldots, C_K$ with no two adjacent nodes in the same class) permits full GPU parallelism within each class, with sequential processing across the $K \leq 8$ classes. For a Texas-scale graph ($|V| \approx 914\text{K}$, $|E| \approx 2.7\text{M}$), the storage requirements are modest by modern GPU standards. Actual iteration counts and sweep times depend on implementation, memory layout, move set, and annealing schedule, and should be reported as part of numerical experiments rather than treated as theoretical guarantees.

**Correspondence table.**

| Continuum (Path 1) | Discrete |
| --- | --- |
| Arc $\Gamma_{ij}$, curvature $\kappa_{ij}(s) = (P_i - P_j)d(s)$ | Boundary edges $(u, v)$ with $f(u) = i$, $f(v) = j$; $\Delta E_{\text{eff}} = 0$ at equilibrium |
| Arc length $\mathcal{H}^1(\Gamma_{ij})$ | $\sum_{f(u)=i,\, f(v)=j} w_{uv}$ |
| Density $d(s)$ on arc | Block population $p_v$ |
| Pressure $P_i - P_j$ | District pressure $\alpha(\text{pop}_i - P/N)$ |
| Density-modified VPMCF velocity $v_n = \kappa + (P_j - P_i)d(s)$ | Block reassignment rate $\propto \Delta E_{\text{eff}}$ |
| $120^\circ$ junction (continuum) | Locally stable three-way meeting (discrete, $\pm 10^\circ$) |

---

# 5. A Gerrymandering Metric

## 5.1 Definition and Decomposition

Let $L_{\text{exist}}$ be the boundary length of the existing map, $L_{\text{local}}$ the boundary length at the nearest local minimum (obtained by gradient descent from the existing map), and $L_{\text{global}}$ the boundary length of the global minimizer.

**Definition 5.1.**
$\text{GM}_{\text{tac}} = \frac{L_{\text{exist}}}{L_{\text{local}}} \geq 1, \qquad \text{GM}_{\text{sys}} = \frac{L_{\text{local}}}{L_{\text{global}}} \geq 1, \qquad \text{GM}_{\text{total}} = \frac{L_{\text{exist}}}{L_{\text{global}}} \geq 1.$

The decomposition is multiplicative ($\text{GM}_{\text{total}} = \text{GM}_{\text{tac}} \times \text{GM}_{\text{sys}}$) and additive in logarithms. (Additionally, the logarithm representation accurately captures the soap-bubble method's characterization of "zero gerrymandering," namely when $L_{\text{exist}}$ = $L_{\text{global}}$, $log(\text{GM}_{\text{total}}) = 0$.) Being quotients of like kind, the GM variables are dimensionless, and so may be validly used to compare different states' districting plans.

$\text{GM}_{\text{tac}} > 1$ captures **boundary-level manipulation**: the existing map has unnecessarily jagged boundaries that could be smoothed, without redrawing which areas belong together, to reduce $L$. $\text{GM}_{\text{sys}} > 1$ captures **structural manipulation**: the fundamental grouping of communities is itself geometrically costly, and no amount of boundary smoothing can correct it.

## 5.2 Approximating the Global Minimum

$L_{\text{global}}$ is approximated by running simulated annealing from $M$ random initializations and taking $\hat{L}_{\text{global}} = \min_m L^{(m)}$.

**LP lower bound.** A lower bound on $L_{\text{global}}$ can in principle be obtained by solving an LP relaxation of the integer assignment problem: replacing binary block-assignment variables with continuous $[0, 1]$ variables and dropping the contiguity constraint, at census tract resolution. This relaxation is tractable with standard solvers. However, fractional assignments permit "boundary" contributions to be spread across many partially-assigned tracts, potentially producing a lower bound that lies well below $L_{\text{global}}$ and is therefore of limited practical use as a certificate. The LP bound is best regarded as a rough order-of-magnitude check rather than a tight guarantee. Tightening it requires either (a) adding integrality constraints on a subset of tracts via branch-and-bound, feasible for small states, or (b) penalizing fractional assignments directly in the objective.

**Primary quality certificate.** For large states the convergence plot of $\hat{L}_{\text{global}}$ against $M$ is the more reliable diagnostic: a clear plateau in the best-of-$M$ curve indicates that additional restarts are unlikely to improve the estimate. Reporting $\sigma_M = \text{std}(L^{(1)}, \ldots, L^{(M)})$ alongside $\hat{L}_{\text{global}}$ provides a practical measure of approximation stability.

## 5.3 Population-Only Design

The formulation uses total block population and block geometry only. No racial, ethnic, partisan, or other demographic data enters the computation. The scores measure geometric efficiency of district boundaries relative to the equal-population optimum, and are by construction agnostic about the composition of the population being partitioned.

# 6. Discussion

## 6.1 Non-Uniqueness

Problem 2.2 generically admits multiple local minimizers, from topological multiplicity (distinct adjacency structures for which districts border which), geometric multiplicity within a fixed adjacency structure, and domain symmetries. The global minimum is not known to be unique for $N \geq 3$. For $N = 2$ in a convex domain, the global minimum is a single geodesic and is generically unique [M94].

Non-uniqueness affects the systemic score: $\mathrm{GM}_{\mathrm{sys}}$ compares the nearest local minimum to the best global approximation. Monitoring convergence of $\hat{L}_{\mathrm{global}}$ with $M$ (Section 5.2) quantifies the uncertainty in this component.

## 6.2 Relation to the Double Bubble Conjecture

The $N = 2$ equal-population case in a circular domain with uniform density is the two-dimensional double bubble problem, proved in [F93]. This provides an exact benchmark for algorithm validation: the algorithm should converge to three mutually tangent circular arcs meeting at $120^\circ$, with $L_{\mathrm{global}}$ agreeing with the analytically computed optimum. For non-uniform density, the corresponding equilibrium arcs satisfy $\kappa(s) = (P_1 - P_2)d(s)$, and no exact solution is known in general, making numerical approximation the only available approach.

## 6.3 Numerical Validation Benchmarks

The preceding analysis suggests several natural validation tests for an implementation.

1. **Uniform-density disk.** For $N = 2$, the optimizer should recover the planar double-bubble geometry: three circular arcs meeting at $120^\circ$. This provides a benchmark against a known analytic solution.
2. **Synthetic urban-rural density.** For a smooth density field such as a Gaussian population center on a rural background, optimized interfaces should satisfy $\kappa_{ij}(s)/d(s) \approx \mathrm{constant}$ along each interface after smoothing at the grid scale.
3. **Resolution study.** Running the same synthetic problem on successively refined grids should reveal whether the discrete Potts equilibria approach the continuum first-order conditions and how strongly grid-pinning affects the result.
4. **Real-state case studies.** States with contrasting geography, such as Iowa, Colorado, Michigan, and Florida, provide useful tests of the treatment of fixed boundaries, water features, and density heterogeneity.

These experiments are not needed to define the variational model, but they are important for establishing the practical reliability of the proposed gerrymandering metric.

## 6.4 Open Problems

*(i) Convergence of density-modified VPMCF.* Long-time convergence of $v_n = \kappa + (P_j - P_i)d(s)$ for $N \geq 3$ districts with variable $d$ and topology changes is not fully established. Phase-field regularizations provide a well-posed substitute [ESS92].

*(ii) Quantifying the gap between Path 1 and Path 2 optima.* Bounding $|L(\mathcal{P}_L^*) - L(\mathcal{P}_{\tilde{L}}^*)|$ in terms of the Lipschitz constant of $\log d$ would clarify when the two objectives produce substantially different maps, particularly relevant for states with large urban-rural density contrasts.

*(iii) Grid-pinning correction.* Establishing a formal convergence rate for the discrete algorithm (with cluster swaps) to the density-modified VPMCF fixed points, accounting for the grid-pinning effects of irregular block geometries, would place the computational approach on a rigorous footing.

*(iv) Approximation ratio.* No polynomial-time algorithm is known guaranteeing a constant-factor approximation to the global minimum of Problem 2.2 with the contiguity constraint. The approximation quality of best-of-$M$ annealing as a function of $N$ and domain geometry is open.

*(v) Uniqueness conditions.* Conditions on $\Omega$ and $d$ under which the global minimum of Problem 2.2 is unique remain open for $N \geq 3$.

---